\documentclass[sigconf]{acmart}
\usepackage{amsmath}
\usepackage{filecontents}
\date{}
\usepackage{times}
\usepackage{algorithm}
\usepackage{tikz}
\usepackage{amsthm,amsmath,amssymb,array,color,colortbl,graphicx,multirow}
\usepackage{comment}
\usepackage{balance}
\usepackage{xspace}
\usepackage{wrapfig}
\usepackage{enumitem}
\usepackage{makecell}

\definecolor{Gray}{gray}{0}
\definecolor{RED}{RGB}{255,0,0}
\definecolor{GREEN}{RGB}{0,255,0}
\definecolor{BLUE}{RGB}{0,0,255}
\definecolor{YELLOW}{RGB}{255,255,0}
\definecolor{BLACK}{RGB}{0,0,0}

\def\M{\mathcal{M}}

\definecolor{applegreen}{rgb}{0.55, 0.71, 0.0}

\date{}

\usepackage{blindtext}
\usepackage{color}
\usepackage{url}
\usepackage{xspace}
\usepackage{amssymb} 
\usepackage{amsmath} 
\usepackage{amsthm}

\usepackage{mathtools}
\usepackage{float}
\usepackage{graphicx}
\usepackage{footnote}
\usepackage{comment}
\makesavenoteenv{tabular}
\makesavenoteenv{table}
\usepackage{multirow}
\usepackage{booktabs}
\usepackage[flushleft]{threeparttable}
\usepackage{tikz}
\usetikzlibrary{calc}
\usepackage{pgfplots}
    \usepgfplotslibrary{colorbrewer}
    \pgfplotsset{
        cycle list/Set1-9,
        cycle multiindex* list={
            mark list*\nextlist
            Set1-9\nextlist
        },
    }
    \usepgfplotslibrary{fillbetween}
    \usetikzlibrary{patterns}
\usepackage{algpseudocode}
\usepackage[acronym]{glossaries}
\glsdisablehyper
\usepackage[utf8]{inputenc}
\usepackage[english]{babel}
\usepackage{hyperref}
\hypersetup{
 colorlinks=true, 
 linkcolor=blue,
 filecolor=magenta,      
 urlcolor=cyan,
}

\usepackage[font=small]{caption}

\usepackage{subcaption}
\newcommand{\rot}{RotorNet\xspace}

\usepackage{amsfonts}

\renewcommand{\ss}{\textsc{Static}\xspace}
\newcommand{\rs}{\textsc{Rotor}\xspace}
\newcommand{\cs}{\textsc{Demand-aware}\xspace}

\newcommand{\longPL}[1]{\ignorespaces}

\newcommand{\model}{\textsc{TMT}\xspace}

\usepackage{subcaption}
  
\def\compactify{\itemsep=0pt \topsep=0pt \partopsep=0pt \parsep=0pt}
\let\latexusecounter=\usecounter

\usepackage{color}
\definecolor{lightgray}{gray}{0.75}

\usepackage[noframe]{showframe}
\usepackage{framed}
\usepackage{lipsum}
 {\endMakeFramed}
\definecolor{shadecolor}{gray}{0.75}

\newcommand{\netw}{N}
\newcommand{\netws}{\mathcal{N}}

\newcommand{\Cost}{\mathrm{Cost}}
\newcommand{\RecCost}{\mathrm{adj}}
\newcommand{\RouCost}{\mathrm{srv}}

\newcommand{\card}[1]{\lvert #1\rvert}
\def\A{\mathcal{A}}

\renewcommand\footnotetextcopyrightpermission[1]{} 
\setcopyright{none}
\renewcommand\footnotetextcopyrightpermission[1]{} 
\renewcommand\@formatdoi[1]{\ignorespaces}
\makeatother

\begin{document}


\title[ToR-to-ToR Matching Model]{An Online Matching Model for\\Self-Adjusting ToR-to-ToR Networks}

\author{Chen Avin}
\affiliation{
	\institution{Ben Gurion University, Israel}
}
\email{avin@cse.bgu.ac.il }

\author{Chen Griner}
\affiliation{
	\institution{Ben Gurion University, Israel}
}
\email{griner@post.bgu.ac.il}

\author{Iosif Salem}
\affiliation{
	\institution{University of Vienna, Austria}
}
\email{iosif.salem@univie.ac.at}

\author{Stefan Schmid}
\affiliation{
	\institution{University of Vienna, Austria}
}
\email{stefan_schmid@univie.ac.at}
\begin{abstract}
    This is a short note that formally presents the matching model for the theoretical study of self-adjusting networks as initially proposed in \cite{avin2019toward}.
\end{abstract}

\maketitle
\setcopyright{none}

\section{Background and Motivation}
\label{sec:motivation}

This note is motivated by the observation that  existing 
datacenter network designs sometimes provide a \emph{mismatch} 
between some common traffic patterns and the switching technology used 
in the network topology to serve it.  On the contrary, we make the case for a systematic 
approach to assign a specific type of traffic or flow to 
the topology component which best matches its characteristics and requirements.
For instance, static topology components 
can provide a very low latency,
however, static topologies inherently require  multi-hop forwarding: 
the more hops a flow has to traverse, the more network capacity is consumed, 
which can be seen as an ``bandwidth tax,'' as noticed in prior work~\cite{rotornet}. 
This makes these networks less fitted at high loads: 
the more traffic they carry, the more bandwidth tax is paid. Inspired by the notion of bandwidth tax, 
we introduce a second dimension, called ``latency tax'' to capture the delay incurred by the reconfiguration time of 
optical switches. For instance, rotor switches reduce the bandwidth tax by providing periodic \emph{direct} connectivity. 
While this architecture performs well for all-to-all traffic patterns, it is less suited for elephant flows created by ring-reduce traffic pattern of machine learning training with Horovod. We note that static and rotor topology components both form 
demand-oblivious topologies, and hence, they cannot 
account for specific elephant flows. While 
Valiant routing \cite{valiant1982scheme} can be used in combination with rotor switches
to carry large flows, this again results in bandwidth tax.
This is the advantage of demand-aware topologies, based on 3D MEMS optical circuit switches, 
which can provide shortcuts specifically to such elephant flows. However, the state-of-the-art demand-aware optical switches 
have a reconfiguration latency of several milliseconds and hence incur a higher latency tax to establish a circuit. Moreover, demand-aware topologies might require a control logic that adds to the latency tax. Thus, this latency can only be amortized for large flows, which benefit from the demand-aware topology components in the longer term.

\begin{figure}[t!]
\begin{centering}
\includegraphics[width=\columnwidth]{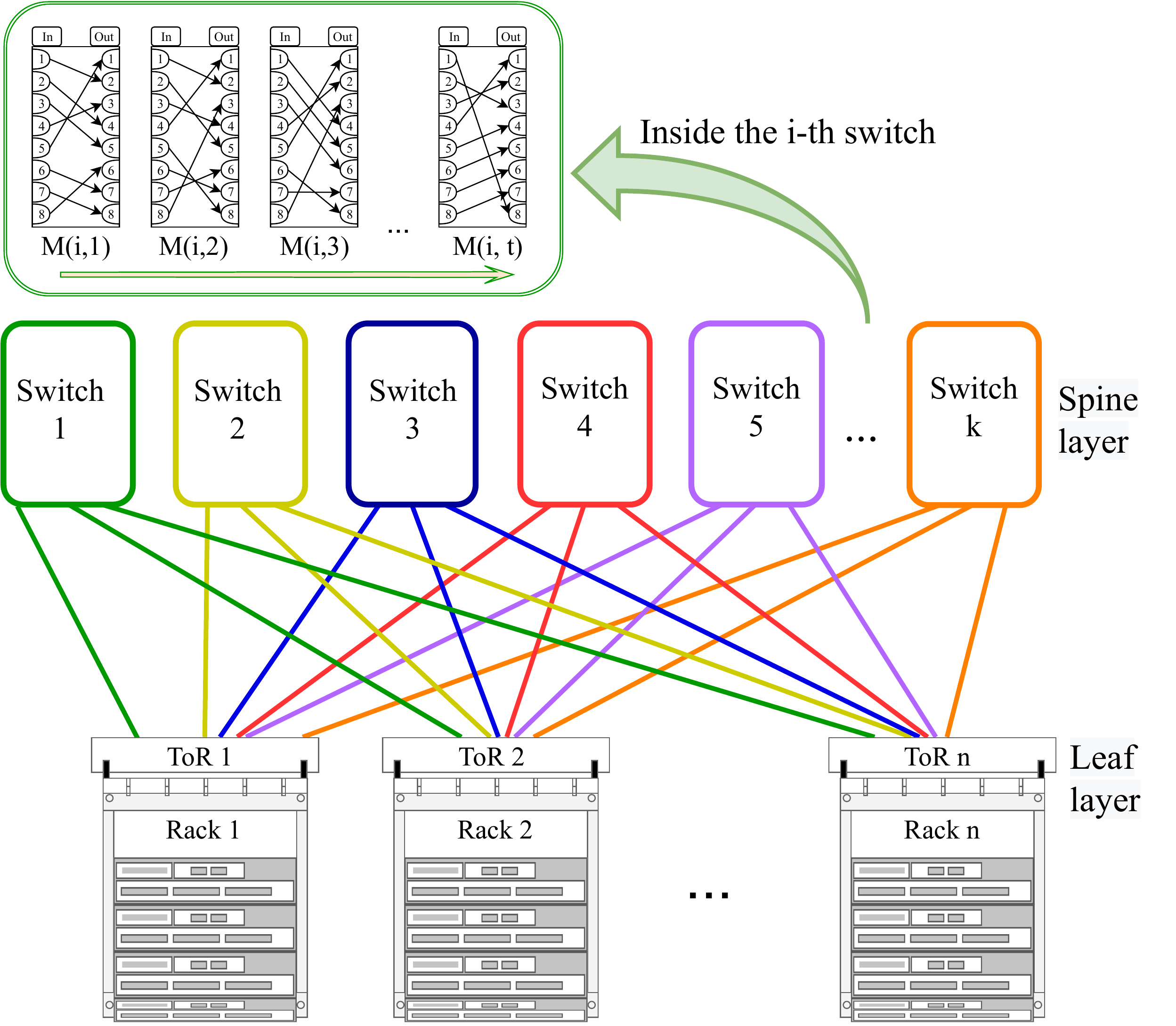}     
\caption{Overview of TMT model.}
\label{fig:system}
\end{centering}
\end{figure}

\section{ToR-Matching-ToR Architecture}

Given the above motivation for a unified network design,
combining the advantages of static, rotor, and demand-aware switches, 
we propose a two-layer leaf-spine network architecture in which 
spine switches can be of different types 
\ss, \rs and \cs.
Since this network architecture generalizes existing
architectures such as RotorNet \cite{rotornet},
in that it supports different types of switches 
\emph{matching} ToRs to each other, we will refer to it as the 
ToR-Matching-ToR (\model) 
network model.

More specifically, the \model  network 
interconnects a set of $n$ ToRs, $\{1,2, \dots, n\}$
and its two-layer leaf-spine architecture composed of
 leaf switches and spine switches, similar  to~\cite{opera,rotornet}. 
The $n$ ToR packet switches are connected using $k$ spine  switches, 
$SW = \{sw_1, sw_2, \dots, sw_k\}$ and each switch internally connects its 
in-out ports via a matching. Figure~\ref{fig:system} illustrates 
a schematic view of our design. We assume that each ToR $i: 1 \le i \le n$ has 
$k$ uplinks, where uplink $j: 1 \le j \le k$  connects to port $i$ in $sw_j$.  
The directed outgoing (leaf) uplink is connected to incoming port of the (spine) switch 
and the directed incoming (leaf) uplink is connected to the outgoing port of the (spine) switch. 
Each switch has $n$ input ports and $n$ output ports and the connections are directed, 
from input to output ports. 

At any point in time, each switch  $sw \in SW$ provides a \emph{matching} between its input 
and output ports. Depending on the switch type, this matching may be \emph{reconfigured} at runtime: 
The set of matchings $\M_j$ of a switch $j$ may be larger than one, i.e., $m_j=\card{\mathcal{M}_j}>1$. 
Changing from a matching $M' \in \mathcal{M}$ to a matching $M''\in \mathcal{M}$ takes time, which we model 
with a parameter $R_j$:  the \emph{reconfiguration time} of switch $j$. During reconfiguration, the links in $M' \cap M''$, i.e., the links which are not being reconfigured, can still be used for forwarding; the remaining links are blocked during  the reconfiguration. Depending on the technology, different switches in $SW$ support different sets of matchings and reconfiguration times.

We note that the TMT network can be used to model
existing systems, e.g.,
Eclipse \cite{venkatakrishnan2018costly} 
or ProjecToR \cite{projector} which
rely on a demand-aware switches, \rot \cite{rotornet} 
and Opera \cite{opera} 
which relies on a rotor based switches, 
or an optical variant of Xpander~\cite{xpander} 
which can be built from a collection of static matchings.

\section{The Matching Model}

This section presents a general algorithmic model 
for Self-Adjusting Networks (SAN) constructed using a set of matchings. 
We mostly follow \cite{avin2019toward}.
We consider a set of~$n$ nodes $V=\{1,\ldots,n\}$
(e.g., the top-of-rack switches). 
The communication \emph{demand} among these nodes
is a sequence 
$\sigma =
(\sigma_1, \sigma_2, \ldots)$ of \emph{communication
requests} where~$\sigma_t = (u,v)  \in V \times V$, 
is a source-destination pair. 
The communication demand can either
be finite or infinite.

In order to serve this demand, the nodes~$V$ must
be inter-connected by a network~$\netw$, defined over the 
same set of nodes. In case of a demand-aware network,
$\netw$ can be optimized towards~$\sigma$,
either statically or dynamically:
a self-adjusting network~$\netw$ can change over time, and we denote
  by~$\netw_t$ the network at time~$t$, i.e.,
	the network evolves: $\netw_0,$ $\netw_1,$ $\netw_2,$ $\ldots$ 

\subsection{Matching}

The $n$ nodes are connected using $k$ switches, 
$SW = \{sw_1, sw_2, \dots, sw_k\}$ and each switch internally connects its $n$ in-out ports via a \emph{matching}. 
These matchings can be dynamic, and change over time. To denote the matching on a switch $i$ at time $t$ we use $M(i,t)$. At each time $t$ our network is the union of these matchings,  $\netw_t=\mathcal{M}=\bigcup_{i=1}^{k} M(i,t)$. 

In general, not all switches are necessarily reconfigurable. Since, reconfigurable switches tend to be more costly than static ones, a network could gain from using some hybrid mix of switches. 

\subsection{Cost}

The crux of designing smart self-adjusting networks is 
to find an optimal \emph{tradeoff} between the benefits
and the costs of reconfiguration:
while by reconfiguring the network, we may be able to serve
requests more efficiently in the future, reconfiguration
itself comes at a cost.

The inputs to the \emph{matching based} self-adjusting network design problem 
is the number of nodes $n$, the number of switches (i.e., matchings) $k$,  
a set of allowed network topologies $\netws$ (i.e., all networks that can be built from $k$ matchings),
the request sequence $\sigma=(\sigma_1,\sigma_1,\ldots,
\sigma_{m})$, and two types of costs:
\begin{itemize}
\item An \textbf{adjustment cost} $\RecCost: \netws \times \netws \rightarrow \mathbb{R}$ which defines the cost of reconfiguring a network
 $\netw \in \netws$ to a network  $\netw' \in \netws$. Adjustment  costs may include
 mechanical costs (e.g., energy required to move lasers or abrasion) as well as 
 performance costs  (e.g., reconfiguring a network may entail control plane
 overheads or packet reorderings,
 which can harm throughput).  
 For example, the cost could be given 
by the number
of links which need to be changed in order to transform
the network. 
\item A \textbf{service cost} $\RouCost: \sigma \times
\netws \rightarrow \mathbb{R}$ which defines,
for each request $\sigma_i$ and for each network $\netw\in \netws$,
what is the price of serving $\sigma_i$ in network $\netw$.
For example, the cost could correspond to the route length:
shorter routes require less resources and hence reduce not only
load (e.g., bandwidth consumed along fewer links),
but also energy consumption, delay, and flow completion
times, could be considered for example.
\end{itemize}

Serving request $\sigma_i$ under the current
network configuration $\netw_{i}$
will hence cost $\RouCost(\sigma_i,\netw_i)$,
after which the network reconfiguration algorithm may decide
to reconfigure the network at cost $\RecCost(\netw_{i},\netw_{i+1})$. 
The total processing cost of a demand sequence $\sigma$
for an algorithm $\A$
is then
\begin{align}
\Cost(\A, \netw_0, \sigma) = \sum_{t=1}^{m} \RouCost(\sigma_t,\netw_{t-1}) 
+ \RecCost(\netw_{t-1},\netw_{t}) 
\end{align}
\noindent where $\netw_t \in \mathcal{N}$ denotes the network at time $i$.

\subsection{Specific Metrics}

\subsubsection{Service Cost}

In order to give a more useful description of the performance of a self-adjusting network,
we model the service cost for each $\sigma_i=(u,v)$ as the shortest distance between $u$ and $v$ on the graph $N_i$, that is $$\RouCost(\sigma_t,\netw_{t})=d_{\netw_t}(u,v)=d(\sigma_t,\netw_t)$$
\noindent where $d_{G}(u,v)$ denotes the \emph{shortest path} distance between $u$ and $v$ on the graph $G$.

\subsubsection{Adjustments Cost}

Adjustments cost can depend on the particular network modeled.  
We will discuss three particular cases for adjustment costs and recall that our network graph at time $t$, $N_t$, 
is a union of the different matchings on each of our $k$ switches. 
At any time $t$, a switch can adjust its matching, causing a change in the overall network's topology. 
\begin{itemize}
    \item \textbf{Edge Distance:} The basic case where we define the adjustments cost 
    as propositional to the number of replaced edges between each consecutive matchings of the same switch. 
    Recall that we denote the matching of switch $i$ at time $t$ as $M(i, t)$, 
    which denotes the set edges in the the matching. 
    Let the cost of a single edge be $\alpha$ then
    the adjustment cost for a single switch is therefore $\alpha \cdot \card{M(i, t+1) \setminus M(i, t)}$, 
    where $S \setminus T$ denotes the \emph{set difference} between $S$ and $T$.
    For the entire network, this turns out to be 
    $$\RecCost(\netw_{t-1},\netw_{t})=\alpha \sum_{i=1}^k \card{M(i,t)\setminus M(i,t-1)}.$$ 
    
    \item \textbf{Switch Cost:} In this case, if a matching (switch) is changed, 
    it costs $\alpha$ regardless of the number of edge changes in the matching. 
    Let $\mathbb{I}_{S=T}$ be an indicator function that denotes if set $S$ is equal set $T$. 
    Then the adjustments cost for the network is:
    $$\RecCost(\netw_{t-1},\netw_{t})=\alpha \sum_{i=1}^k \mathbb{I}_{M(i,t) = M(i,t-1)}.$$
    
    \item \textbf{No Direct Cost:} In this case the adjustment cost is zero
    $$\RecCost(\netw_{t-1},\netw_{t})=0$$
    however the cost of reconfiguring the network is still incurred 
    through the inactivity of some of the edges during the adjustment itself. 
    When some switch $i$ changes its matching from $M(i,t)$ to $M(i,t+1)$, 
    its edges will be unavailable, and requests cannot be served using these edges 
    until the adjustment process is completed after some $\beta$ units of time. 
    Here, we also consider two cases: (i) the entire switch (matching) is unavailable for $\beta$ time units, 
    namely all its edges are inactive; (ii) only the edges that are changing are inactive for $\beta$ time units. 
    Let $M^*(i,t)$ denote the set of \emph{active} edges in matching $M(i,t)$ (or in $sw_i$).
    Then for each time $t$ we have:
    $$\netw_t=\bigcup_{i=1}^{k} M^*(i, t)$$
    
\end{itemize}

{\balance
  \bibliographystyle{ieeetr} 
   \balance
\bibliography{bibs/literature,bibs/cerberus,bibs/bibly}
}
\end{document}